\documentclass[aps,prl,twocolumn,superscriptaddress]{revtex4}
\usepackage{graphicx}
\usepackage{latexsym}
\usepackage{amssymb}
\usepackage{amsmath}
\usepackage{amsfonts}
\usepackage{bm}
\usepackage{multirow}
\usepackage{color}

\newcommand{\dsZ}{\mathbb{Z}}
\newcommand{\ii}{\mathrm{i}}
\newcommand{\dd}{\mathrm{d}}

\newcommand{\vect}[1]{{\bm{#1}}}
\newcommand{\eqnref}[1]{Eq.\,\eqref{#1}}
\newcommand{\figref}[1]{Fig.\,\ref{#1}}

\newcommand{\beq}{\begin{equation}}
\newcommand{\eeq}{\end{equation}}
\newcommand{\beqn}{\begin{eqnarray}}
\newcommand{\eeqn}{\end{eqnarray}}

\begin{document}

\title{Anyon and Loop Braiding Statistics in Field Theories with a Topological $\Theta-$term}

\author{Zhen Bi}

\affiliation{Department of physics, University of California,
Santa Barbara, CA 93106, USA}

\author{Yi-Zhuang You}

\affiliation{Department of physics, University of California,
Santa Barbara, CA 93106, USA}

\author{Cenke Xu}

\affiliation{Department of physics, University of California,
Santa Barbara, CA 93106, USA}

\begin{abstract}

We demonstrate that the anyon statistics and three-loop statistics
of various $2d$ and $3d$ topological phases can be derived using
semiclassical nonlinear Sigma model field theories with a
topological $\Theta$-term. In our formalism, the braiding
statistics has a natural geometric meaning: The braiding process
of anyons or loops leads to a nontrivial field configuration in
the space-time, which will contribute a braiding phase factor due
to the $\Theta$-term.


\end{abstract}

\pacs{}

\maketitle

\emph{Introduction} ---

One of the key properties of topological states is that, the
gapped topological excitations above the ground state can have
nontrivial braiding statistics. In both $2d$ and $3d$, all
discrete lattice gauge theories have a deconfined topological
phase~\cite{polyakovbook}. $2d$ discrete gauge theories have point
particle topological excitations, while $3d$ discrete gauge
theories have both particle excitations and loop excitations which
correspond to gauge charge and gauge flux loop respectively. The
simplest lattice discrete gauge theory (which we call ``plain
gauge theory") already has nontrivial braiding
statistics~\cite{kitaev2003}. More exotic gauge theories can be
constructed by coupling the plain gauge theory to matter fields,
and drive the matter fields into certain nontrivial short range
entangled (SRE) state or symmetry protected topological (SPT)
phase~\cite{wenspt,wenspt2}. For example, once we couple a $2d$
$p+\ii p$ topological superconductor to a $\dsZ_2$ gauge field,
then the vison of the gauge field would acquire a Majorana fermion
zero mode, which will grant the vison a nonabelian
statistics~\cite{ivanov,e8}. Also, if we couple a $2d$ bosonic SPT
phase with $\dsZ_2$ symmetry to a $\dsZ_2$ lattice gauge theory,
the lattice gauge theory will have both semion and anti-semion
excitations~\cite{levingu}, which is different from a plain
lattice gauge theory.

\begin{figure}[htbp]
\begin{center}
\includegraphics[width=240pt]{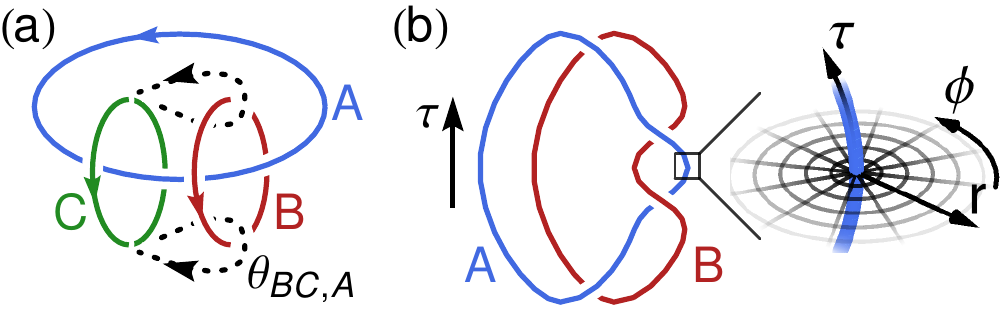}
\caption{(Color online.) (a) Three-loop braiding process. The
loops A, B and C are colored blue, red and green respectively. The
braiding path of loop-C is indicated by the dotted arrow curve.
(b) Two-loop linking in the $(2+1)d$ space-time, which corresponds
to creating a pair of $\dsZ_2^A$ and $\dsZ_2^B$ visons, and
annihilating them after braiding one $\dsZ_2^A$ and one $\dsZ_2^B$
visons. The time $\tau$ is along the vertical direction. The inset
shows the local cylindrical coordinate system around a segment of
the $\dsZ_2^A$ vison loop.} \label{fig: loops}
\end{center}
\end{figure}

Recently these results have been generalized to $3d$ systems. It
was demonstrated that once a $3d$ lattice discrete gauge theory is
coupled to a $3d$ SPT state, the loop excitations (fluctuating
gauge flux loops) would acquire nontrivial multi-loop braiding
statistics~\cite{levinloop,wenloop1,wenloop2,ranloop}, in addition
to the standard particle-loop statistics of the plain gauge
theory. For example when loop-B and loop-C are both linked to
loop-A, namely none of the loops is contractible, the system wave
function could acquire a universal phase angle after braiding
loop-C through loop-B as shown in \figref{fig: loops}(a). These
braiding statistics can be used as a diagnostics for SPT
phases~\cite{levinloop}.

Besides the standard group cohomology description of SPT phases
introduced in Ref.~\onlinecite{wenspt,wenspt2}, it was pointed out
in Ref.~\onlinecite{senthilashvin,xu3dspt,xusenthil,xuclass} that
the bosonic SPT phases can also be described by semiclassical
nonlinear sigma model (NLSM) field theories with a topological
$\Theta$-term. In this theory all the field variables are
fluctuating Landau order parameters that transform nontrivially
under global symmetry. The goal of this work is to demonstrate
that the nontrivial statistics between topological excitations
after coupling the SPT phases to a discrete gauge theory can also
be described and calculated using this NLSM field theory.
Basically the braiding phase factor comes from the $\Theta-$term
in the field theory, as long as we carefully analyze the field
configuration in the space-time which corresponds to the braiding
process. The NLSM field theory with a topological term can be
viewed as the continuum limit field theory description for these
braiding statistics.

\emph{2d Anyon statistics} ---

We will first look at $2d$ systems, and as an example let us start
with the $2d$ SPT state with $\dsZ_2^A \times \dsZ_2^B$ symmetry,
which can be described by the following $(2+1)d$ O(4) NLSM with a
$\Theta$-term at $\Theta = 2\pi$~\cite{xuclass}:
\begin{equation} S =\int \dd^2x\dd\tau \ \frac{1}{g}
(\partial_\mu \vect{n})^2 + \frac{\ii\Theta}{
\Omega_3}\epsilon_{abcd} n^a \partial_x n^b
\partial_y n^c \partial_\tau n^d, \label{o4nlsm}
\end{equation}
where $\vect{n}$ is a four component vector with unit length, and
$\Omega_3=2\pi^2$ is the volume of a three dimensional sphere with
unit radius. Under the $\dsZ_2^A \times \dsZ_2^B$ symmetry, the
vector $\vect{n}$ transforms as \beqn \dsZ_2^A: n^1, n^2
\rightarrow - n^1, -n^2, \ \ n^3, n^4 \rightarrow n^3, n^4; \cr\cr
\dsZ_2^B: n^1, n^2 \rightarrow n^1, n^2, \ \ n^3, n^4 \rightarrow
- n^3, - n^4. \eeqn Now let us couple the vector $\vect{n}$ to a
$\dsZ_2^A \times \dsZ_2^B$ gauge field. The excitations that will
have nontrivial braiding statistics are the vison excitations
($\pi$-gauge flux) of gauge fields $\dsZ_2^A$ and $\dsZ_2^B$. Let
us consider the following braiding process: one pair of $\dsZ_2^A$
visons and one pair of $\dsZ_2^B$ visons are created in space at
one instance in time, then they are annihilated at another later
instance after braiding one $\dsZ_2^A$ vison with one $\dsZ_2^B$
vison. In the $(2+1)d$ space-time, this process corresponds to one
linking between $\dsZ_2^A$ and $\dsZ_2^B$ vison loops, as shown in
\figref{fig: loops}(b). Because the $\dsZ_2$ gauge fields are
coupled to the four-component vector $\vect{n}$, the $\dsZ_2^A$
vison is bound with a $\pm 1/2$-vortex of $(n^1, n^2)$, while
$\dsZ_2^B$ vison is bound with a $\pm 1/2$-vortex of $(n^3, n^4)$.
Then the braiding process in the space-time can be viewed as a
linking configuration between $(n^1, n^2)$ half-vortex loop and
$(n^3, n^4)$ half-vortex loop. Due to the $\Theta$-term in
\eqnref{o4nlsm}, this configuration will contribute a phase factor
$\exp(\pm \ii \pi /2) = \pm \ii$ to the action, which implies the
mutual braiding statistics between the $\dsZ_2^A$ vison and
$\dsZ_2^B$ vison.

To calculate this phase factor explicitly, let us first consider a
finite segment of $\dsZ_2^A$ vison loop along the $\hat{\tau}$
direction. A vison is always bound with either $1/2$-vortex or
$-1/2$-vortex of $(n_1, n_2)$.
Around this segment, the O(4) vector $\vect{n}$ has the following
configuration with cylindrical coordinate $(r, \phi,\tau)$ ($x = r
\cos\phi$, $y = r \sin\phi$, see \figref{fig: loops}(b) inset):
\begin{equation}
\begin{split}
n^1 &=\sin\alpha(r) \cos f(\phi),\\
n^2 &=\sin\alpha(r) \sin f(\phi), \\
n^3 &=\cos\alpha(r)N^1(\tau), \\
n^4 &=\cos\alpha(r)N^2(\tau),
\end{split}
\end{equation} where $\vect{N}=(N_1,N_2)$ is an O(2) unit vector
$|\vect{N}|^2 = 1$. $\vect{N}$ is a function of $\tau$ only.
$\alpha(r)$ is a nonnegative continuous function that satisfies
$\alpha(0) = 0$, $\alpha(\infty) = \pi/2$. Along the $\hat{\tau}$
axis, $i.e.$ $r = 0$, we have $(n^3, n^4) = \vect{N}$. Using this
configuration, we can compute the $\Theta$-term:
\begin{equation}
\begin{split}
&\int \dd^2x \dd\tau \ \frac{2\pi\ii}{ \Omega_3
} \epsilon_{abcd} n^a \partial_x n^b \partial_y n^c \partial_\tau
n^d \\
=&\int_0^{2\pi}\dd\phi\,\partial_\phi f \int \dd\tau\,\frac{\ii}{2\pi}
\epsilon_{ab}N^a
\partial_\tau N^b .
\end{split}
\label{thetareduce1}
\end{equation}
If $n^1$ and $n^2$ form a full vortex line along the $\hat{\tau}$
axis, namely $f(\phi)\sim \phi$, the O(4) $\Theta$-term reduces to
a $1d$ O(2) NLSM with $\Theta = 2\pi$. If there is a $\dsZ_2^A$
vison line along the $\hat{\tau}$ axis, $i.e.$ $n^1$ and $n^2$
form a $\pm1/2$-vortex line along $\hat{\tau}$ axis, namely
$f(\phi) \sim \pm \phi/2$, then the $(2+1)d$ O(4) NLSM reduces to
a $1d$ O(2) NLSM of vector $\vect{N}$ with $\Theta = \pm \pi$. Now
let us consider two linked vison loops, and in
\eqnref{thetareduce1} $\tau$ becomes the parameter along the
$\dsZ_2^A$ vison loop. Since the two loops are linked, vector
$\vect{N}$ will have a $\pm 1/2$-vortex winding along $\dsZ_2^A$
vison loop: \beqn \oint \dd\tau \ \epsilon_{ab}N^a
\partial_\tau N^b = \pm \pi. \label{thetareduce2} \eeqn Combining
\eqnref{thetareduce1} and \eqnref{thetareduce2} together, we
conclude that this linking configuration (which corresponds to a
braiding process in the space-time) would contribute factor $\pm
\ii$ to the action. In other words, the linking configuration in
\figref{fig: loops}(b) corresponds to $\pm 1/4$-instanton of the
four component vector $\vect{n}$ in the $(2+1)d$ space-time.

Now let us consider a $2d$ SPT state with $\dsZ_2$ global symmetry
only, and couple it to a $\dsZ_2$ gauge field. This SPT state can
be described by the same field theory \eqnref{o4nlsm}, and under
the $\dsZ_2$ symmetry $\vect{n} \rightarrow -\vect{n}$. A vison of
this $\dsZ_2$ gauge field can be viewed as a bound state between
the $\dsZ_2^A$ vison and $\dsZ_2^B$ vison discussed previously.
Then the linking configuration in \figref{fig: loops}(b) can be
interpreted as creating a pair of visons, self-twisting one vison
by $2\pi$, then annihilating them. The phase $\pm \ii$ corresponds
to topological spin-$\pm 1/4$ of the vison, which is consistent
with the semion and anti-semion statistics of the vison proved in
Ref.~\onlinecite{levingu}.

All the analysis above can be straightforwardly generalized to
$\dsZ_N$ gauge theory coupled to a $2d$ $\dsZ_N$ SPT state. The
$2d$ $\dsZ_N$ SPT state is described by the same field theory
\eqnref{o4nlsm}~\cite{xuclass}, where $\Theta = 2\pi k$, $k = 0,
1, \cdots (N - 1)$. The same analysis above leads to the result
that the topological spin of the $2\pi/N$ flux excitations can be
$k/N^2$, namely self-twisting such excitation will grant its wave
function a phase $\exp(2\pi \ii k/ N^2)$.

\emph{3d loop statistics} ---

Now we consider $3d$ bosonic SPT states with $\dsZ_2^A \times
\dsZ_2^B \times \dsZ_2^C$ symmetry. In terms of field theory, one
of these SPT states is described by the following $(3+1)d$ O(5)
NLSM: \beqn S = \int \dd^3x \dd\tau \ \frac{1}{g} (\partial_\mu
\vect{n})^2 + \frac{\ii \Theta}{ \Omega_4 } \epsilon_{abcde} n^a
\partial_x n^b \partial_y n^c \partial_z n^d \partial_\tau n^e,
\label{o5nlsm} \eeqn where $\Omega_4=8\pi^2/3$ is the volume of a four
dimensional sphere with unit radius. Under the $\dsZ_2^A \times
\dsZ_2^B \times \dsZ_2^C$ symmetry, the five component vector
$\vect{n}$ transforms as
\begin{equation}
\begin{split}
 \dsZ_2^A:&\; n^1, n^2 \rightarrow -n^1, -n^2, \quad
n^{3,4,5} \rightarrow n^{3,4,5}; \\
 \dsZ_2^B:&\; n^2, n^3
\rightarrow - n^2, - n^3, \quad n^{1,4,5} \rightarrow n^{1,4,5};\\
 \dsZ_2^C:&\; n^4, n^5 \rightarrow - n^4, - n^5, \quad n^{1,2,3}
\rightarrow n^{1,2,3}.
\end{split}
\end{equation}
Now let us couple this SPT state to $\dsZ_2^A \times \dsZ_2^B
\times \dsZ_2^C$ gauge field, and consider the statistics between
the three loops in \figref{fig: loops}(a), in which the base loop
is a vison loop of $\dsZ_{2}^A$ gauge field, and it is linked with
vison loops of both $\dsZ_2^B$ and $\dsZ_2^C$ gauge fields.

\begin{figure}[htbp]
\begin{center}
\includegraphics[width=210pt]{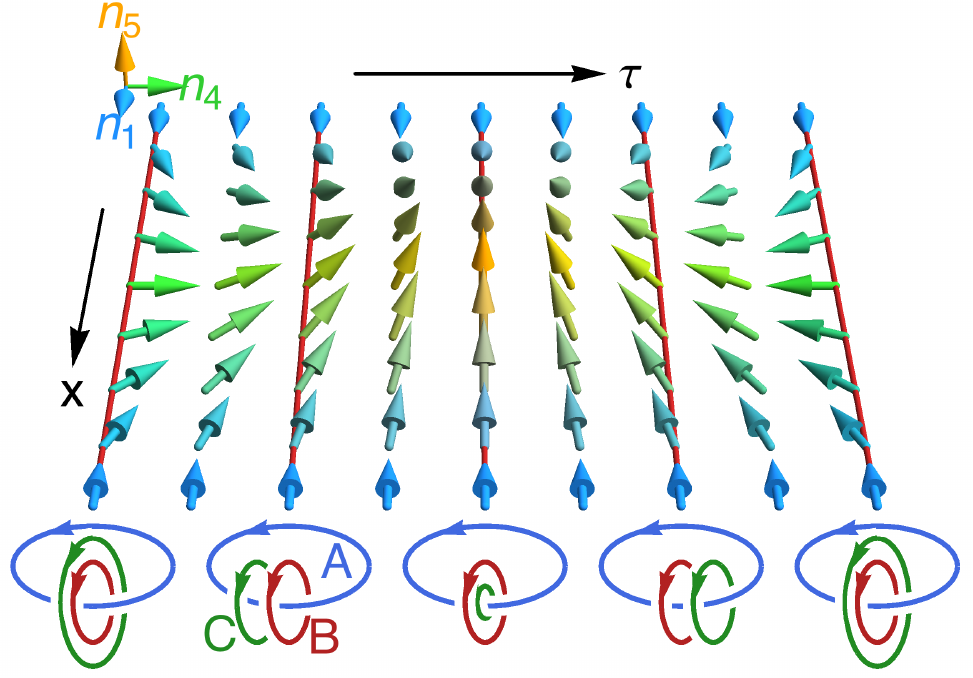}
\caption{(Color online.) The space-time configuration of
$\vect{N}\sim(n_1,n_4,n_5)$ on the world sheet of the $\dsZ_2^B$
vison loop (in red) as the $\dsZ_2^C$ vison loop (in green)
braiding around it. Each red line is a time slice, at which moment
the corresponding three-loop configuration is shown below.}
\label{fig: instanton}
\end{center}
\end{figure}

A vison loop can be bound with either a $+1/2$-vortex or
$-1/2$-vortex, both cases exist in the system, and they correspond
to different excitations. As an example let us study the braiding
statistics of vison loops bound with $+1/2$-vortex. The choice of
$+1/2$ vortex gives each vison loop an orientation, as marked out
in \figref{fig: loops}(a). Let us first look at the $\dsZ_2^B$
vison loop. Following the same calculation as
\eqnref{thetareduce1}, because $\dsZ_2^B$ vison loop is bound with
a half-vortex loop of $(n_2, n_3)$, the O(5) NLSM with $\Theta =
2\pi$ is reduced to an O(3) NLSM with $\Theta = \pi$ in the
$(1+1)d$ world-sheet of the $\dsZ_2^B$ vison loop, and the three
component vector on this world sheet is $\vect{N} \sim (n_1, n_4,
n_5)$: \beqn S_{1d,B} = \int \dd x \dd\tau \
\frac{1}{g}(\partial_\mu \vect{N})^2 + \frac{\ii \pi}{4\pi}
\epsilon_{abc} N^a \partial_x N^b
\partial_\tau N^c. \eeqn On the $(1+1)d$ world sheet of
$\dsZ_2^B$ vison loop, the braiding between $\dsZ_2^B$ and
$\dsZ_2^C$ vison loops corresponds to the space-time configuration
$\vect{N}(x,\tau)$ in \figref{fig: instanton}, and this
configuration carries $1/2$ O(3) instanton number, thus it will
contribute a factor $\ii$ to the action. This implies that the
three-loop braiding statistics angle is $\theta_{BC,A} = \pi/2$.
The statistics angle $\theta_{AC,B}$ can be calculated in the same
way after interchanging $n_1$ and $n_3$ in the O(5) vector, which
will lead to factor $-1$ due to the antisymmetrization in the
$\Theta$-term in \eqnref{o5nlsm}. Thus $\theta_{AC,B} = -\pi/2$.

\begin{figure}[htbp]
\begin{center}
\includegraphics[width=250pt]{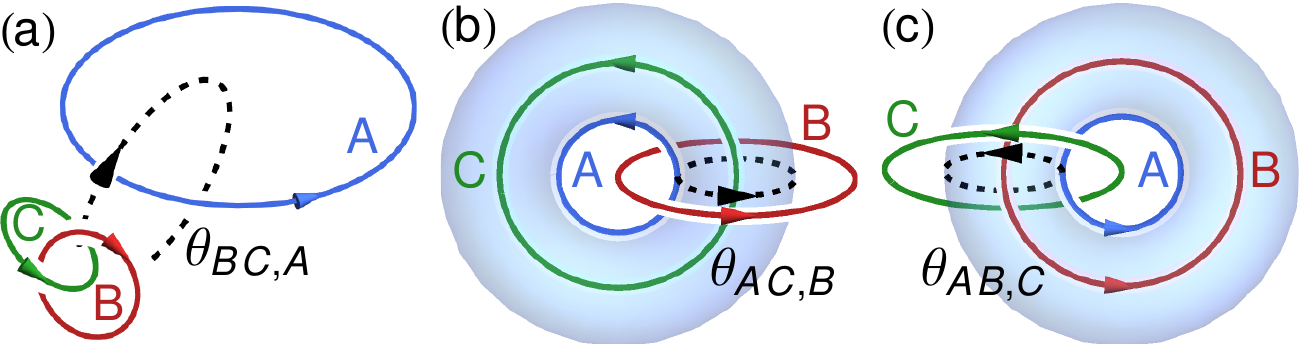}
\caption{(Color online.) (a) Braiding a link of the $\dsZ_2^B$ and
$\dsZ_2^C$ vison loops with the $\dsZ_2^A$ vison loop also
accumulates the phase $\theta_{BC,A}$. (b,c) The three-loop
braiding process that corresponds to the statistic angle
$\theta_{AC,B}$ ($\theta_{AB,C}$). The light blue torus indicates
the surface traced out by the $\dsZ_2^A$ vison loop through the
braiding processes, which can be considered as the Gaussian
surface that measures the $\dsZ_2^A$ charge enclosed. Small arrows
on the loops mark out the loop orientation.} \label{fig: cyclic}
\end{center}
\end{figure}

The loop braiding statistics can also be understood in a different
way. Ref.\,\onlinecite{jianqi} pointed out that the three-loop
braiding in \figref{fig: loops}(a) can also be viewed as a link of
the $\dsZ_2^B$ and $\dsZ_2^C$ vison loops braiding with the
$\dsZ_2^A$ vison loop, as illustrated in \figref{fig: cyclic}(a).
This link-loop braiding statistics can be described by the NLSM as
well. As the vison link braid through the vison loop, the
space-time configuration of the O(5) vector $\vect{n}$ around the
vison link can be described as following:
\begin{equation}\label{eq: O5 vector}
\begin{split}
n^1 &=\cos\alpha(\tau),\\
n^2 &=\sin\alpha(\tau)N^1(x,y,z), \\
n^3 &=\sin\alpha(\tau)N^2(x,y,z), \\
n^4 &=\sin\alpha(\tau)N^3(x,y,z), \\
n^5 &=\sin\alpha(\tau)N^4(x,y,z),
\end{split}
\end{equation} where
$\vect{N}=(N^1,N^2,N^3,N^4)$ is an O(4) unit vector
$|\vect{N}|^2=1$ that describes the configuration of the (linked)
half-vortex loops bound to the vison loops of $\dsZ_2^B$ and
$\dsZ_2^C$. The time $\tau$ (running from 0 to 1) parameterizes a
full braiding of the $\dsZ_2^B\times\dsZ_2^C$ vison link with the
$\dsZ_2^A$ vison loop. Suppose the $n^1$ component is
energetically more favored, then the $\dsZ_2^A$ branch cut disk
bordered by the $\dsZ_2^A$ vison loop will be bound with a $n^1$
domain wall. Let the braiding of the $\dsZ_2^B\times\dsZ_2^C$
vison link initiates from one side of the domain wall, and ends up
at the other side of the domain wall, then $\alpha(\tau)$ will be
a continuous function satisfying $\alpha(0)=\pi$, $\alpha(1)=0$.
Plugging the configuration \eqnref{eq: O5 vector} into the NLSM
\eqnref{o5nlsm}, the O(5) $\Theta$-term of $\vect{n}$ is reduced
to an O(4) $\Theta$-term of $\vect{N}$ at $\Theta=2\pi$: \beqn
\label{eq: O4 of N} &-& \int_0^1\dd\tau\,\partial_\tau
\alpha\sin^3\alpha\int\dd^3x\,\frac{2\pi\ii}{\Omega_4}
\epsilon_{abcd}N^a\partial_x N^b \partial_y N^c \partial_z N^d
\cr\cr &=&
\int\dd^3x\,\frac{2\pi\ii}{\Omega_3}\epsilon_{abcd}N^a\partial_x
N^b \partial_y N^c \partial_z N^d. \eeqn According to our previous
calculation, the linking configuration between $(N_1,N_2)$
half-vortex loop and $(N_3,N_4)$ half-vortex loop corresponds to
the 1/4 O(4) soliton in the $3d$ space, so the above O(4)
$\Theta$-term in \eqnref{eq: O4 of N} will result in a $\pi/2$
phase angle accumulated in the link-loop braiding, which equals to
the three-loop braiding angle $\theta_{BC,A}$ calculated already
in our paper.

The non-trivial link-loop braiding statistics implies that the
$\dsZ_2^B\times\dsZ_2^C$ vison link must carry the charge of the
$\dsZ_2^A$ gauge field. Let us denote the $\dsZ_2^A$ charge
carried by the $\dsZ_2^B\times\dsZ_2^C$ vison link as $q_{BC}^A$.
It is related to the braiding angle by $\theta_{BC,A}=-\pi
q_{BC}^A$. The minus sign is due to the reversed link-loop braiding
direction as shown in \figref{fig: cyclic}(a) (which corresponds
to the positive three-loop braiding direction).
As shown in \figref{fig: cyclic}(b), the torus traced out by the
$\dsZ_2^A$ vison loop through braiding with the $\dsZ_2^C$ vison
loop (in the linking with the $\dsZ_2^B$ vison loop) actually
forms a Gaussian surface enclosing the $\dsZ_2^C$ vison loop. So
the three-loop braiding statistics angle $\theta_{AC,B}$ measures
the $\dsZ_2^A$ charge carried by the $\dsZ_2^C$ vison loop in the
$\dsZ_2^B\times\dsZ_2^C$ link, denoted $q_{C}^A$, and
$\theta_{AC,B}=\pi q_C^A$. Similarly from \figref{fig: cyclic}(c),
the three-loop braiding statistics angle $\theta_{AB,C}$ measures
the $\dsZ_2^A$ charge carried by the $\dsZ_2^B$ vison loop in the
same $\dsZ_2^B\times\dsZ_2^C$ link, denoted $q_{B}^A$, and
$\theta_{AB,C}=\pi q_B^A$. Obviously, $q_{BC}^A=q_B^A+q_C^A$, thus
\beqn \theta_{AB,C}+\theta_{BC,A} + \theta_{AC,B} =0, \eeqn which
is precisely the cyclic relation~\cite{jianqi, levinloop}, and it
implies that $\theta_{AB,C} = 0$ (given $\theta_{BC,A}=\pi/2$ and
$\theta_{AC,B}=-\pi/2$ as previously calculated).

\begin{figure}[htbp]
\begin{center}
\includegraphics[width=190pt]{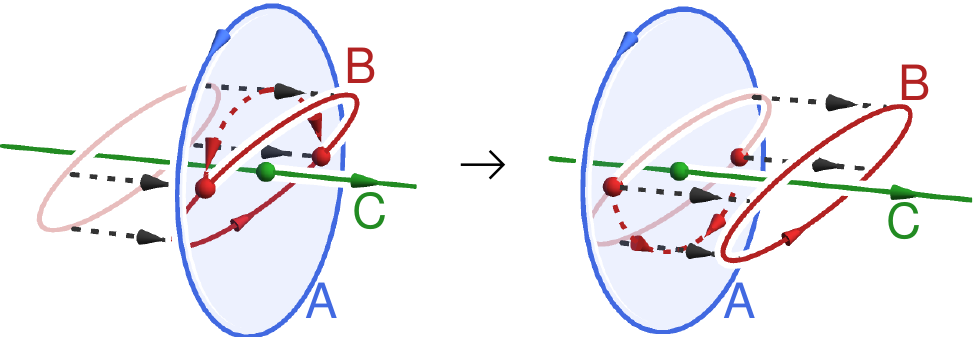}
\caption{(Color online.) Illustration of moving $\dsZ_2^B$ vison
loop through the $\dsZ_2^A$ vison loop. The $\dsZ_2^A$ vison loop
borders a branch cut disk, which can be viewed as a $2d$
$\dsZ_2^B\times\dsZ_2^C$ SPT. When the $\dsZ_2^B$ vison loop pokes
through this disk, a pair of $\dsZ_2^B$ semion-antisemion are
created, braided with the $\dsZ_2^C$ vison, and annihilated.}
\label{fig: braiding}
\end{center}
\end{figure}

$\theta_{AB,C}$ can also be computed as follows: $\theta_{AB,C}$
corresponds to braiding $\dsZ_2^A$ and $\dsZ_2^B$ vison loops,
both of which are linked to a $\dsZ_2^C$ vison loop. This process
can be divided into two steps: first moving $\dsZ_2^B$ vison loop
through $\dsZ_2^A$ vison loop, then moving $\dsZ_2^A$ vison loop
through $\dsZ_2^B$ vison loop. The first step (see \figref{fig:
braiding}) is equivalent to creating a pair of $\dsZ_2^B$
vison-antivison (vison and antivison have semion and antisemion
statistics) at the $2d$ $\dsZ_2^B \times \dsZ_2^C$ SPT phase, then
braiding the $\dsZ_2^B$ vison (or antivison) around the $\dsZ_2^C$
vison, and annihilating the vison-antivison pair. This step will
contribute a phase factor $\ii$ to the action. The second step is
equivalent to creating and annihilating a pair of $\dsZ_2^A$
visons at the $2d$ $\dsZ_2^A \times \dsZ_2^C$ SPT phase, and
braiding around the $\dsZ_2^C$ vison in between, which will
contribute factor $- \ii$. The two processes together will lead to
a trivial phase factor, namely $\theta_{AB,C} = 0$.

More ``conventionally", $\theta_{BC,A}$ and $\theta_{AC,B}$ can be
interpreted in the ``decorated domain wall"
picture~\cite{chenluashvin}. In our NLSM \eqnref{o5nlsm}, the
$\dsZ_2^A$ vison loop is the boundary of a $2d$ disk of branch cut
of coupling between $n_1$ components. According to
Ref.~\onlinecite{xuvisonline}, after integrating out $n_1$, the
effective field theory on this $2d$ disk is the same as
\eqnref{o4nlsm} with $\Theta = 2\pi$, except now the O(4) vector
is $(n_2, n_3, n_4, n_5)$, $i.e.$ this $2d$ disk can be viewed as
a $2d$ SPT state with $\dsZ_2^B \times \dsZ_2^C$ symmetry, which
is precisely the decorated domain wall picture. Then after gauging
the $\dsZ_2^B$ and $\dsZ_2^C$ symmetry, the vison loop statistics
reduces to the anyon statistics of the $2d$ $\dsZ_2^B \times
\dsZ_2^C$ topological order, which is what we have already
computed using \eqnref{o4nlsm}.

We can also consider $3d$ SPT state with $\dsZ_2^A \times
\dsZ_2^B$ symmetry. There are in total three different nontrivial
$3d$ bosonic SPT states with this symmetry~\cite{wenspt}. The
first state can be constructed using the previously discussed
$\dsZ_2^A \times \dsZ_2^B \times \dsZ_2^C$ SPT state, and break
its subgroup $\dsZ_2^B \times \dsZ_2^C$ down to one diagonal
$\dsZ_2$ symmetry, namely now the O(5) vector $\vect{n}$
transforms as
\begin{equation}
\begin{split}
\dsZ_2^A:&\; n_1, n_2
\rightarrow -n_1, -n_2, \quad n_{3,4,5} \rightarrow n_{3,4,5};\\
 \dsZ_2^B:&\; n_1 \rightarrow n_1, \quad  n_{2,3,4,5} \rightarrow
- n_{2,3,4,5}.
\end{split}
\end{equation}
Now a $\dsZ_2^B$ vison loop corresponds to a bound state between
the $\dsZ_2^C$ and $\dsZ_2^B$ vison loops in the previous case.
Thus~\footnote{Here $\theta_{BB,A}$ stands for the full braiding
statistics angle between two $\dsZ_2^B$ vison loops while they are
both linked with a $\dsZ_2^A$ vison loop.} \beqn\begin{split}
&\theta_{BB,A} = 2 \theta_{BC,A} = \pi, \\ &\theta_{AB,B} =
\theta_{AC,B} + \theta_{AB,C} = \pm \pi/2. \end{split}\eeqn All
the other braiding angles are zero. The second type of $3d$ SPT
state corresponds to interchanging $\dsZ_2^A$ and $\dsZ_2^B$
symmetries, thus after gauging the symmetries, $\theta_{AB,A} =
\pm \pi/2$, $\theta_{AA,B} = \pi$. The third type of SPT state is
equivalent to the two SPT states discussed above weakly coupled
together, thus \beqn \theta_{AB,A} = \theta_{AB,B} = \pm \pi/2, \
\ \theta_{AA,B} = \theta_{BB,A} = \pi. \eeqn

In summary, we have computed the anyon braiding statistics, and
three-loop statistics of $2d$ and $3d$ topological phases
constructed by coupling plain gauge theories to bosonic SPT
states. Our calculation is based on semiclassical field theories,
and all the braiding phases naturally come from the topological
$\Theta-$term in the field theory.

We acknowledge the enlightening discussion with Chao-Ming Jian and
Meng Cheng. The authors are supported by the the David and Lucile
Packard Foundation and NSF Grant No. DMR-1151208.


\bibliography{loop}

\end{document}